\documentclass[preprint,aps,prd,groupedaddress,showpacs]{revtex4}
\usepackage{graphicx}
\usepackage{amsmath}
\usepackage{bm}
\usepackage{epsfig}

\usepackage{relsize}
\RequirePackage{xspace}

\begin{document}

\title{\mbox{}
Pair Production of Heavy Quarkonium and $B_c(^*)$ Mesons at Hadron
Colliders}

\author{Rong Li$~^{(a)}$, Yu-Jie Zhang$~^{(a)}$, and Kuang-Ta Chao$~^{(a,b)}$}
\affiliation{ {\footnotesize(a)~ Department of Physics and State Key
Laboratory of Nuclear Physics and Technology, Peking University,
 Beijing 100871, China}\\
{\footnotesize (b)~Center for High Energy Physics, Peking
University, Beijing 100871, China}}

\begin{abstract}

We investigate the  pair production of S-wave heavy quarkonium at
the LHC in the color-singlet mechanism (CSM) and estimate the
contribution from the gluon fragmentation process in the color-octet
mechanism (COM) for comparison. With the matrix elements extracted
previously in the leading order calculations, the numerical results
show that the production rates are quite large for the pair
production processes at the LHC. The $p_t$ distribution of double
$J/\psi$ production in the CSM is dominant over that in the COM when
$p_t$ is smaller than about 8GeV. For the production of double
$\Upsilon$, the contribution of the COM is always larger than that
in the CSM. The large differences in the theoretical predictions
between the CSM and COM for the $p_t$ distributions in the large
$p_t$ region are useful in clarifying the effects of COM on the
quarkonium production.   We also investigate the pair production of
S-wave $B_c$ and $B_c^*$ mesons, and the measurement of these
processes is useful to test the CSM and extract the LDMEs for the
$B_c$ and $B_c^*$ mesons.
\end{abstract}

\maketitle

\section{Introduction}
Heavy quarkonium provides a ideal system to investigate both the
perturbative and non-perturbative aspects of quantum chromodynamics.
Conventionally, the color-singlet mechanism (CSM) is used to
describe the decay and production of heavy quarkonium\cite{CSM}. In
the CSM, the processes are factorized into two steps. Firstly the
heavy quark pair are created perturbatively at short distances with
the same color and angular momentum as the final quarkonium state,
and then evolve into the quarkonium non-perturbatively at
long-distances. There have been, however, some problems in the CSM,
e.g., the infrared divergences in the calculation of the decay of
P-wave quanrkonium\cite{BCGR} and the higher-order correction
calculation of S-wave quanrkonium\cite{BGR}, and the surplus
$J/\psi$ production\cite{CDF} of which the rate is much higher than
that of the color-singlet prediction at the Tevatron. The
non-relativistic quantum chromodynamics (NRQCD) factorization
formalism\cite{BBL}, which was put forward by Bodwin, Braaten, and
Lepage, overcame the infrared divergence difficulties in the
color-singlet model\cite{BBL92}, and gave the proper prediction for
the charmonium production at the Tevatron\cite{BFBY}. In  NRQCD, the
heavy quark pair at short distances are not necessarily in the
color-singlet state but can be in the states with different color
and angular-momentum from that of the final state quarkonium. The
color-octet pair can evolve to the color-singlet charmonium by
emitting soft gluons. This is called the color-octet mechanism
(COM).

%
%

Lots of work have been done to investigate the validity and
limitation of the NRQCD formulism in heavy quarkonium production.
The current experimental results on $J/\psi$ photoproduction at HERA
are fairly well  described by the NLO color singlet piece except the
$J/\psi$
polarizations~\cite{Kramer:1995nb,Chang:2009uj,Artoisenet:2009xh}.
The DELPHI data favor the NRQCD color-octet mechanism for $J/\psi$
production $\gamma \gamma \rightarrow J/\psi X$
\cite{Abdallah:2003du,Klasen:2001cu}. The observed large cross
sections of inclusive charmonium production at the Tevatron once
gave strong support to the color-octet gluon fragmentation in NRQCD,
but recently it is found that the NLO results in the color-singlet
piece can bring an order of magnitude enhancement to the $J/\psi$
production rate in the large $p_t$ region\cite{CMT} with
longitudinally polarized $J/\psi$\cite{GW}. The theoretical
prediction for $p_t$ distribution of the $\Upsilon$ production can
properly describe the Tevatron data by including the contributions
from the NLO results and the real correction part at the
next-to-next-to-leading-order in the CSM \cite{ACLMT}. The cross
sections of $J/\psi$ exclusive and inclusive production in $e^+e^-$
annihilation at B factories ~\cite{Abe:2002rb} are much larger than
the LO NRQCD predictions\cite{Braaten:2002fi,cs}, but the
discrepancies seem to be resolved by considering the higher order
effects: NLO QCD corrections\cite{Zhang:2008gp,Ma:2008gq,
Gong:2008ce} and relativistic corrections \cite{He:2007te} without
invoking the color-octet contributions\cite{Ma:2008gq} (discussions
in the light-cone approach can be seen in \cite{bondar}). Recent
developments and related topics in quarkonium production can be
found in
Refs.~\cite{Brambilla:2004wf,Lansberg:2006dh,Lansberg:2008zm}.



The above mentioned developments in heavy quarkonium production
indicate that the situation is far from being conclusive, and
further tests for the color-singlet and color-octet mechanisms in
NRQCD are still needed to clarify various problems involved in heavy
quarkonium production.

In order to investigate the effects of the COM on the production of
heavy quarkonium, it is useful to study processes which heavily
depends on the production mechanism. The pair production of heavy
quarkonium can serve as the desired process.  In NRQCD, the gluon
fragmentation gives the main contribution to the pair production of
quarkonium in the large $p_t$ region of the heavy quarkonium. In the
pair production processes, there appear two long-distance matrix
elements (LDMEs).  So the difference of theoretical predictions
between the CSM and COM could be more obvious. Moreover, because of
charge-parity $C$ conservation the gluon fusion processes $g+g \to
J/\psi+\chi_c$ and $g+g \to J/\psi+\eta_c$ are forbidden in the CSM.
But the gluon fragmentation in the COM can produce these associated
final states. So to detect two final heavy quarkonium states with
different $C$-parity may give a good way to test the COM.

The pair production of heavy quarkonium at hadron colliders has been
studied by many authors. The   color-octet gluon fragmentation into
double charmonium at the Tevatron in NRQCD were considered as
evidence for the COM \cite{BFP}. The CSM prediction on the double
charmonium production was made and it was found that the
contribution with $p_t<4GeV$ in the CSM is dominant\cite{Qiao}. Only
in the large $p_t$ region, the CSM and COM give manifestly different
predictions for the double charmonium production. The Large Hadron
Collider (LHC) is expected to produce a huge number of heavy
quarkonium. Therefore, it is natural to investigate the pair
production of heavy quarkonium at the LHC.

At the LHC, it is also interesting to study the production of the
double heavy flavored mesons $B_c$ and $B_c^*$. The $B_c(^*)$ meson
production in hadron collisions has been studied in
QCD\cite{chang,berezhnoy,baranov}. Some study of $B_c(^*)$ pair
production was also performed in $pp$ and $\gamma\gamma$
collisions\cite{Baranov}. It is useful to extend the study to the
pair production of the $B_c(^*)$ mesons at the LHC.

In this paper, we study the pair productions of heavy
quarkonium at the Tevatron and LHC, including $J/\psi$, $\eta_c$,
$\Upsilon$ and $\eta_b$ in the CSM.
For comparison, the color-octet contributions to the pair
production of $J/\psi$ and $\Upsilon$ are estimated by considering
the gluon fragmentation process.
We also investigate the pair productions of
S-wave $B_c$ and $B_c^*$ mesons where there is no contribution from the
gluon fragmentation process and the COM contributions are
suppressed by the small $v^2$ (the relative velocity between quark
and anti-quark). Therefore, these processes can give
a better test of CSM and be used to extract the
LDMEs of the $B_c$ and $B_c^*$ mesons.

The outline of our paper is as follows. In section 2, some of the
definitions and formulas are given for deriving the cross sections
of the processes. Then the numerical results are presented in
section 3. Finally, in section 4 we give the summary.

\section{The formulations}
\subsection{Color-singlet part}
\begin{figure}[!htbp]
 \begin{center}
  \includegraphics[width=0.90\textwidth]{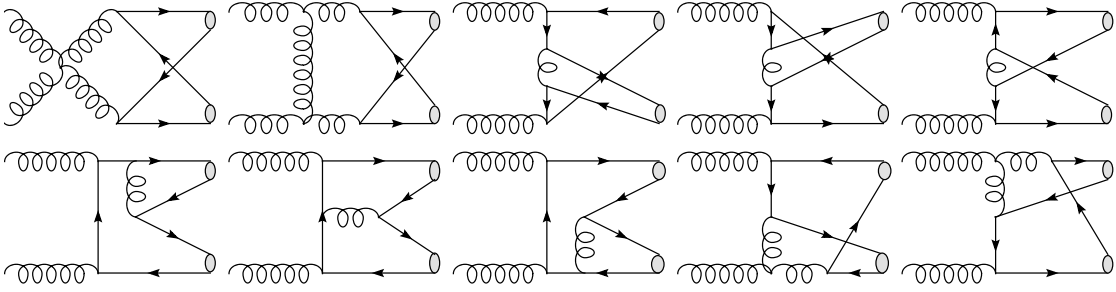}
    \caption{The typical Feynman diagrams for $g+g \to J/\psi+ J/\psi$.
The others can be obtained by reversing the fermion lines or interchanging the initial gluons.
     As for $g+g\to B_c(B_c^*)+\bar{B}_c(\bar{B}_c^*)$, there are same diagrams.}
   \label{fig:diagram1}
 \end{center}
\end{figure}

\begin{figure}[!htbp]
 \begin{center}
  \includegraphics[width=0.40\textwidth]{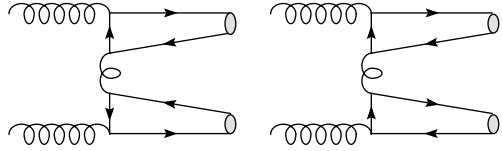}
    \caption{The additional typical Feynman diagrams of $g+g \to \eta_c+
    \eta_c$. The others can be obtained by reversing the fermion
    lines or interchanging the initial gluons.
     }
   \label{fig:diagram2}
 \end{center}
\end{figure}

At hadron collider, the pair production of heavy quarkonium at the
leading-order (LO) in the CSM have two subprocesses $g+g \to {\cal
Q}_1+{\cal Q}_2$ and $q+\bar{q} \to {\cal Q}_1+{\cal Q}_2$. But we
just consider the gluon fusion process in the calculation since it
is the dominant one. There are 31 Feynman diagrams for the
processes $g+g \to J/\psi+ J/\psi$ and $g+g\to
B_c(B_c^*)+\bar{B}_c(\bar{B}_c^*)$. The typical Feynman diagrams are
presented in Fig.~\ref{fig:diagram1}. For the process $g+g \to \eta_c+
\eta_c$, there are additional 8 Feynman diagrams which are showed in
Fig.~\ref{fig:diagram2}. The Feynman diagrams for the process of
$\Upsilon$ and $\eta_b$ production are as same as the corresponding
process of charmonium production.

Following the color-singlet factorization
formalism, the amplitude of the pair production
of S-wave heavy quarkonium is written as
\begin{eqnarray}
\label{CSMderiv3} &&{\cal M}(a+b \rightarrow {\cal Q}_1+{\cal
Q}_2)\nonumber \\&=& \sum_{s_1,s_2,s_3,s_4,i,j,k,l} N_1(\lambda|
s_1,s_2)N_2(\lambda| s_3,s_4) \frac{\delta^{ij}}{\sqrt{N_c}}
\frac{\delta^{kl}}{\sqrt{N_c}} \frac{R_1(0)}{\sqrt{4 \pi}}
\frac{R_2(0)}{\sqrt{4 \pi}} \nonumber \\&& \times {\cal M}(a+b
\rightarrow Q_i \bar Q_j(\mathbf{p_1}=\mathbf{0};s_1,s_2) + Q_k \bar
Q_l(\mathbf{p_2}=\mathbf{0};s_3,s_4))\,
\end{eqnarray}
where the $s_i$ is the spin of the heavy quark in the meson;
$R_i(0)$ is the wave function at
origin of the heavy quarkonium; $\frac{\delta^{ij}}{\sqrt{N_c}}$ is the color project
operator; $N_i(\lambda| s_1,s_2)$ is the spin project operator as
following
\begin{eqnarray}
\label{pojectionfac} N(\lambda| s_1,s_2) = \frac{\sqrt{M_{Q\bar{Q}}}
\epsilon^\lambda_{\mu} \bar{v} (\mathbf{P}_Q,s_2) \gamma^\mu u
(\mathbf{P}_{\bar{Q}},s_1)}{4 m_Q m_{\bar{Q}} },
\end{eqnarray}
where $M_{Q\bar{Q}}$ is the mass of the heavy quarkonium.

We analytically calculate the amplitude square of these subprocesses and present the
analytical formulas of the pair production of $\eta_c(\eta_b)$ and $J/\psi(\Upsilon)$ in
appendix. For $B_c$ and $B_c^*$, the analytical formulas
are too tedious to be presented in this paper. The formula for $d\hat{\sigma}/dt$ on
the double $J/\psi$ production is the same
as that in reference\cite{Qiao}. The final result can be obtained
by convoluting the parton level cross section
with the parton distribution function $f_{g/p}(x)$ as following
\begin{eqnarray}
\label{xs} d\sigma(p+p(\bar{p}) \rightarrow {\cal Q}_1{\cal
Q}_2+X)=\int dx_1 dx_2 f_{g_1/p}(x_1,\mu_f) f_{g_2/p(\bar{p})}(x_2,\mu_f)
d\hat{\sigma}(g_1 + g_2 \rightarrow {\cal Q}_1{\cal Q}_2,\mu_r),
\end{eqnarray}
where the $\mu_r$ and $\mu_f$ are the renormalization and factorization scale.

\subsection{Color-octet Part}

As a comparison, we naively estimate the pair production of $J/\psi$
and $\Upsilon$ in the COM by using the similar way as in
Ref.\cite{BFP} in which the evolution of the fragmentation function
was ignored. In the COM, a gluon can fragment into a $c\bar{c}$ pair
with the quantum number $^3S_1^{(8)}$ and then hadronize
into $J/\psi$. This process will give
large contribution to the double heavy
quarkonium production in the large $p_t$ region and is expressed as
\begin{eqnarray}
\label{fragment1} d\hat{\sigma}_{{\cal Q}_1+{\cal Q}_2} =
\int_0^1dz_1\int_0^1dz_2D_{g \to {\cal Q}_1}(z_1,m_{Q_1})D_{g \to
{\cal Q}_2}(z_2,m_{Q_2})d\hat{\sigma}_{gg}(E_1/z_1,E_2/z_2),
\end{eqnarray}
where $\hat{\sigma}_{gg}$ is the cross section of two real gluon
production; $D$ is the fragmentation function for a gluon to
fragment into a quarkonium. In NRQCD, this fragmentation function is
written as\cite{BBL}
\begin{eqnarray}
\label{function1} D_{g \to {\cal Q}}(z,\mu^2)=\sum_{n}d_{g \to
n}(z,\mu^2){\langle } {\cal O}^{H}_n {\rangle}.
\end{eqnarray}
The short distance coefficient can be calculated perturbatively and
the result of the LO calculation is
\begin{eqnarray}
\label{function2} d_{g \to \underline{8}{}^3S_1}=\frac{\pi
\alpha_s(2m_Q)}{24m^3_Q}\delta(1-z).
\end{eqnarray}
The contribution from $gg(q\bar{q}) \to gg$ subprocesses
is calculated and the contributions from the feeddown of $\psi'$,
$\chi_{cJ}(1P)$, $\Upsilon(2S)$ and $\chi_{bJ}(1P)$ are also
included. The final result is expressed as
\begin{eqnarray}
\label{fragment2} d\hat{\sigma}_{{\cal Q}_1+{\cal Q}_2} &=&
d\hat{\sigma}_{gg}(\frac{\pi\alpha_s(4m_c^2)}{24m_c^3})^2[\langle O_8^{J/\psi}(^3S_1) \rangle +
\langle O_8^{\psi'}(^3S_1) \rangle Br(\psi' \to J/\psi) \nonumber\\&&+
\sum_{J=0}^{2}(2J+1)\langle O_8^{\chi_{c0}(1P)}(^3S_1) \rangle Br(\chi_{cJ} \to J/\psi)]^2.
\end{eqnarray}
Here it is noteworthy that the identical particle factor $``2"$ has been
put in the calculation of $d\hat{\sigma}_{gg}$.

\section{The numerical results and conclusions}

In calculating the numerical results, we choose the following
parameters $M_c=1.5$GeV and $M_b=4.9$GeV, and set the
renormalization and factorization scale as
$\mu_r=\mu_f=\sqrt{4m_Q^2+p_t^2}$. For the gluon fragmentation
process, the renormalization and factorization scale are chosen as
the transverse momentum $p_t(g)$ of the gluon with $p_t(J/\psi)\simeq
p_t(g)$. The parton distribution of CTEQ6L1\cite{CTEQ6} is
used. Therefore, the running of $\alpha_s$ is evaluated by the LO
formula of CTEQ6. The center-of-mass energies of the Tevatron and
LHC are 1.96TeV and 14TeV respectively. The pseudorapidity cuts on
the final quarkonium states are chosen as $-0.6<\eta<0.6$ at the
Tevatron and $-2.4<\eta<2.4$ at the LHC. We use the wave functions
at origin that are calculated by using the logarithmic
potential\cite{EQ,QR} with quark masses almost the same as what we
use. The values for them are listed as:
\begin{eqnarray}
|R(0)|_{c\bar{c}(1S)}^2=0.815\mathrm{GeV^3}, \nonumber \\
|R(0)|_{b\bar{c}(1S)}^2=1.508\mathrm{GeV^3}, \nonumber \\
|R(0)|_{b\bar{b}(1S)}^2=4.916\mathrm{GeV^3}.
\end{eqnarray}

In order to give the results of gluon fragmentation
processes for comparison, the following color-octet matrix
elements are used as the input parameters\cite{9911436,0106120}
\begin{eqnarray}
\langle O_8^{J/\psi}(^3S_1) \rangle=0.39 \times 10^{-2} \mathrm{GeV^3}, &&
\langle O_8^{\Upsilon(1S)}(^3S_1) \rangle=15 \times 10^{-2}\mathrm{GeV^3}, \nonumber \\
\langle O_8^{\psi'}(^3S_1) \rangle=0.37 \times 10^{-2}\mathrm{GeV^3}, &&
\langle O_8^{\Upsilon(2S)}(^3S_1) \rangle=4.5 \times 10^{-2}\mathrm{GeV^3}, \nonumber \\
\langle O_8^{\chi_{c0}(1P)}(^3S_1) \rangle=0.19\times 10^{-2}\mathrm{GeV^3}, &&
\langle O_8^{\chi_{b0}(1P)}(^3S_1) \rangle=4.0\times 10^{-2}\mathrm{GeV^3}.
\end{eqnarray}
These LDMEs are extracted from the matching between the LO NRQCD predictions and the Tevatron
data by using the CTEQ5L parton distribution function with LO $\alpha_s$ running.
From the LO formula of $\alpha_s$ running of CTEQ6, the
corresponding $\alpha_s$ in the fragmentation function for
the $J/\psi$ and $\Upsilon$ are chosen as
$\alpha_s(M_{J/\psi})$=0.286, $\alpha_s(M_{\Upsilon})$=0.201 respectively. The branching
ratios in Eq.~(\ref{fragment2}) are taken from the PDG08\cite{PDG}.
\begin{table}
\caption{The cross sections of pair production of $J/\psi$,
$\Upsilon$, $B_c$ and $B_c^*$ at the Tevatron and LHC with $p_t>$3GeV.}
\label{tab:total}
\begin{center}
\renewcommand{\arraystretch}{1.5}
\[
\begin{array}{|c|c|c|}
\hline\hline
 Final~States & \sigma_{Tevatron}[nb] & \sigma_{LHC}[nb] \\
 \hline
 \eta_c \eta_c  & 3.32\times 10^{-3} & 2.73
 \\[-1mm]
 J/\psi J/\psi& 5.63 \times 10^{-2} & 2.83
 \\[-1mm]
  \eta_b \eta_b & 1.87 \times 10^{-5} & 7.36 \times 10^{-3}
 \\[-1mm]
  \Upsilon \Upsilon & 1.23 \times 10^{-4} & 1.51 \times 10^{-2}
 \\[-1mm]
  B_c\bar{B_c} & 3.86 \times 10^{-3} & 2.72 \times 10^{-1}
 \\[-1mm]
  B_c\bar{B_c^*} & 1.00 \times 10^{-3} & 8.37 \times 10^{-2}
 \\[-1mm]
  B_c^*\bar{B_c^*} & 8.23 \times 10^{-3} & 7.08 \times 10^{-1}
 \\[-1mm]
 \hline \hline
\end{array}
\]
\renewcommand{\arraystretch}{1.0}
\end{center}
\end{table}

In the Table~\ref{tab:total}, we give the cross sections of pair
production of $J/\psi$ ($\Upsilon$) and
$B_c$ ($B_c^*$) at the Tevatron and LHC with $p_t>3$GeV in the
CSM. From the table, the cross section of each process is enhanced by an order or
more in magnitude at the LHC than that at the Tevatron. Therefore,
the LHC will be a good place to study the pair production processes
carefully.

\begin{figure}
\begin{tabular}{cc}
\epsfig{file=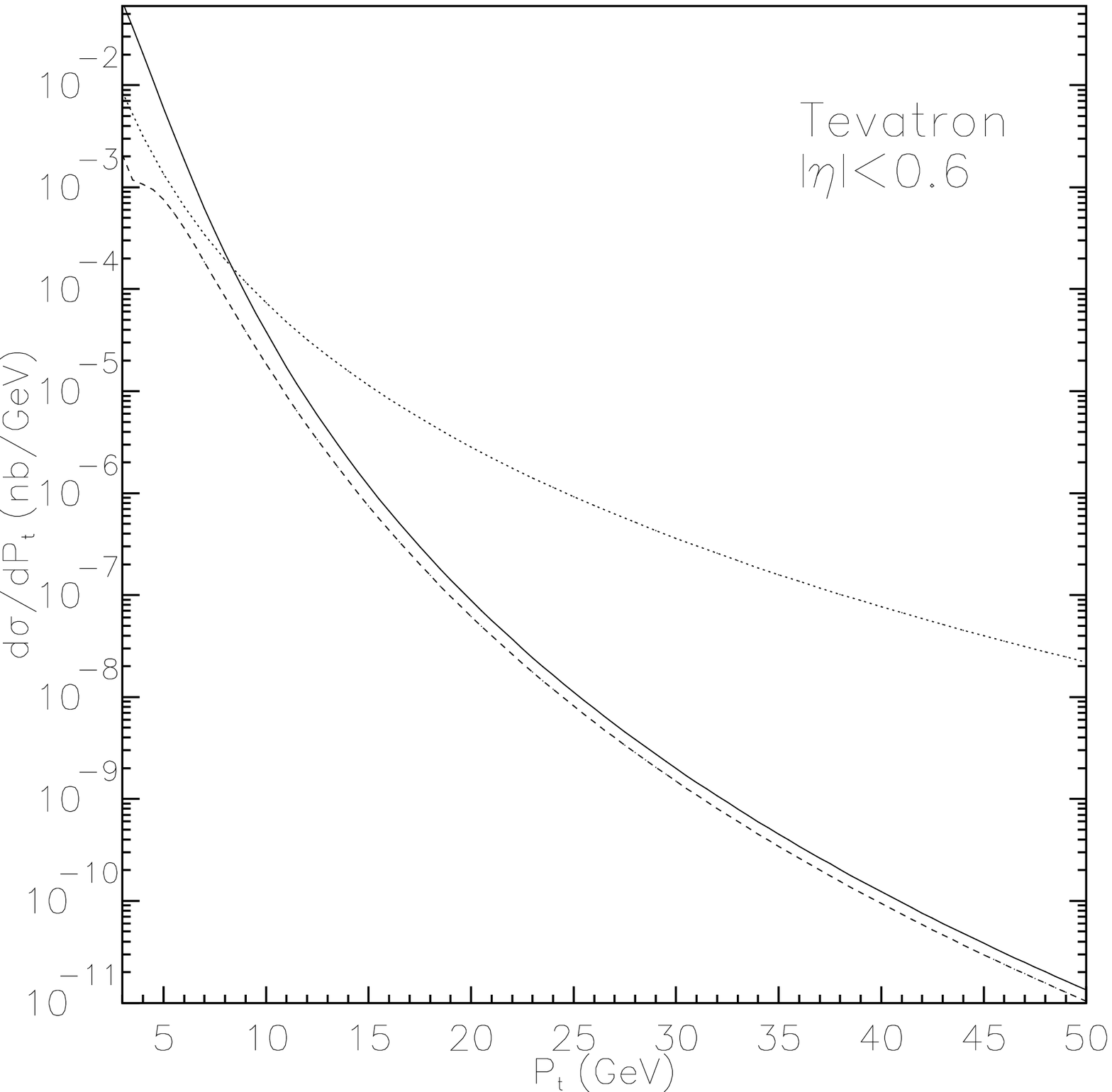,width=8cm}&
\epsfig{file=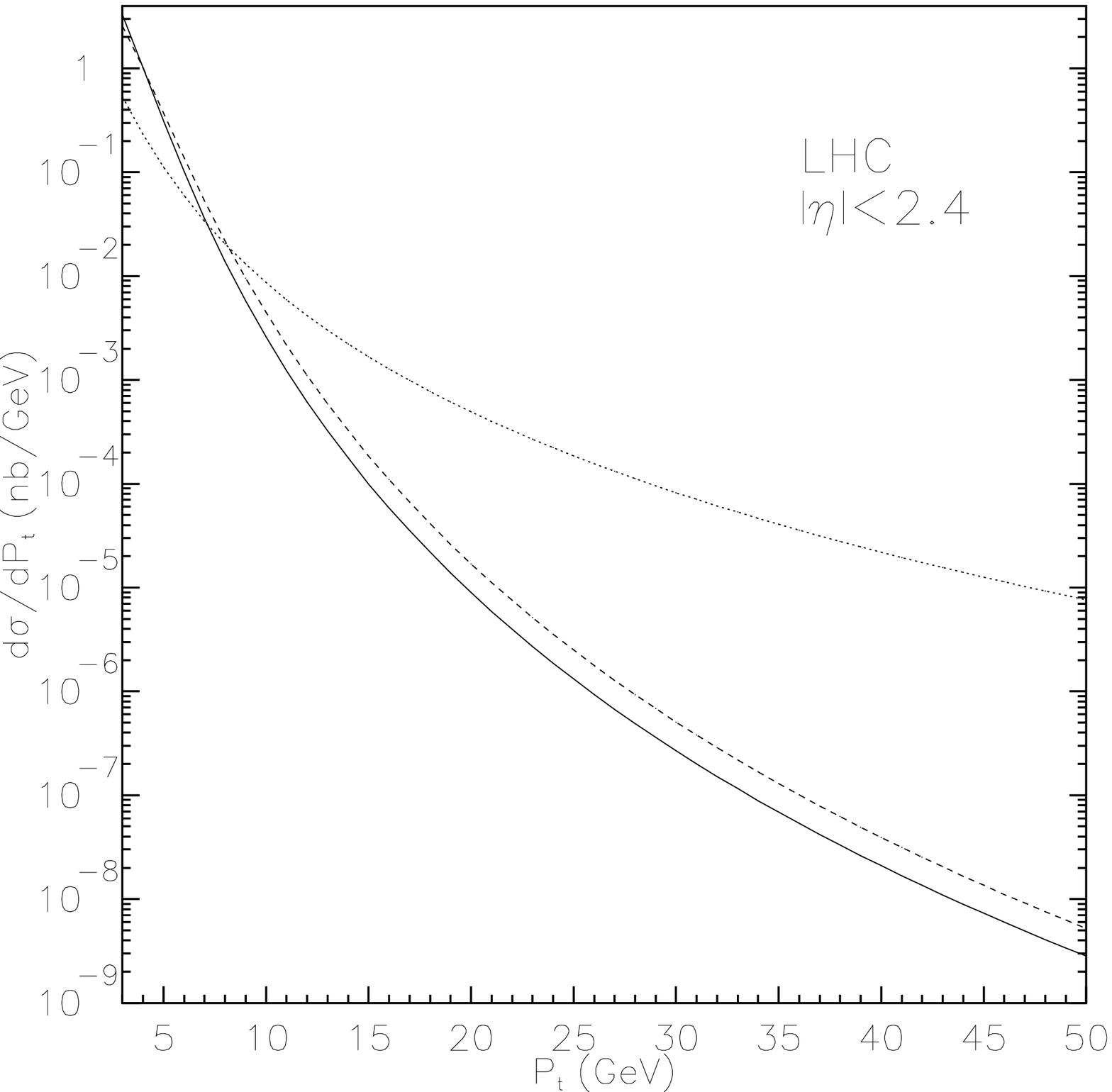,width=8cm}\\
\end{tabular}
\caption{The pair production of $J/\psi$ (solid line) and $\eta_c$ (dashed line)
in the CSM at the hadron
colliders. The dotted line corresponds to the pair production of $J/\psi$ that
come from the gluon fragmentation
process in the COM.}
\label{fig:Jpsi-Etac}
\end{figure}

Fig.~\ref{fig:Jpsi-Etac} shows the $p_t$ distribution of the
pair production for $J/\psi$ and $\eta_c$ at the Tevatron and LHC.
The result of the gluon fragmentation process in the COM is also plotted in the
figure. We can see that whether at the Tevatron or at the LHC the
$p_t$ distributions of $J/\psi$ and $\eta_c$ are similar and the
numerical results at the LHC are enhanced by an order or more in magnitude at
large $p_t$ region than that at the Tevatron. Therefor, the LHC will provide
a chance to measure the $p_t$ distribution of $J/\psi$ pair production.
Comparing the $J/\psi$ pair production in the CSM with that
in the COM, the formal is dominant as the $p_t$ is smaller
than about 8GeV. And the result in the
COM become dominant and even larger
than that in the CSM for three orders in the large $p_t$ region.

\begin{figure}
\begin{tabular}{cc}
\epsfig{file=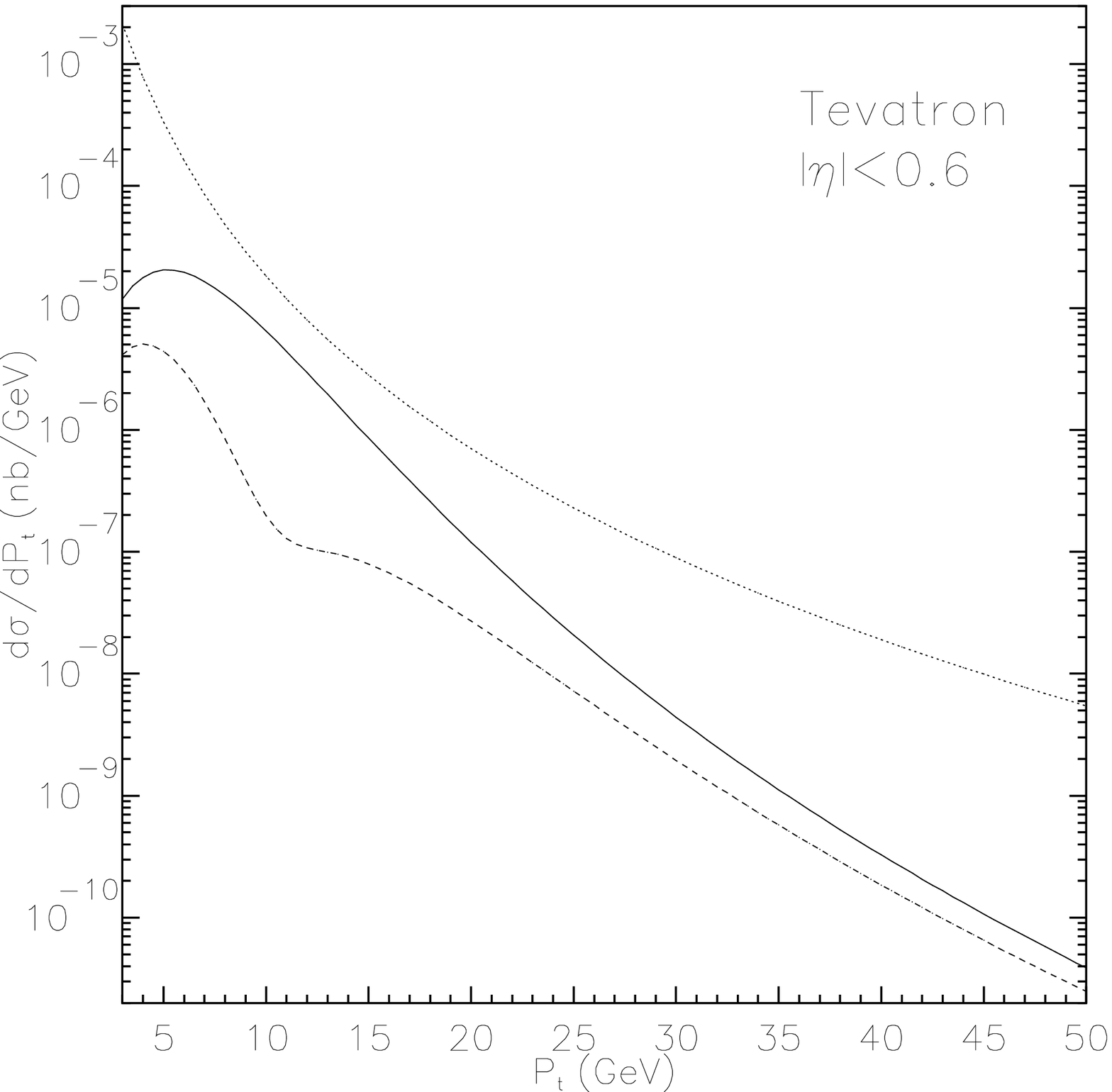,width=8cm}&
\epsfig{file=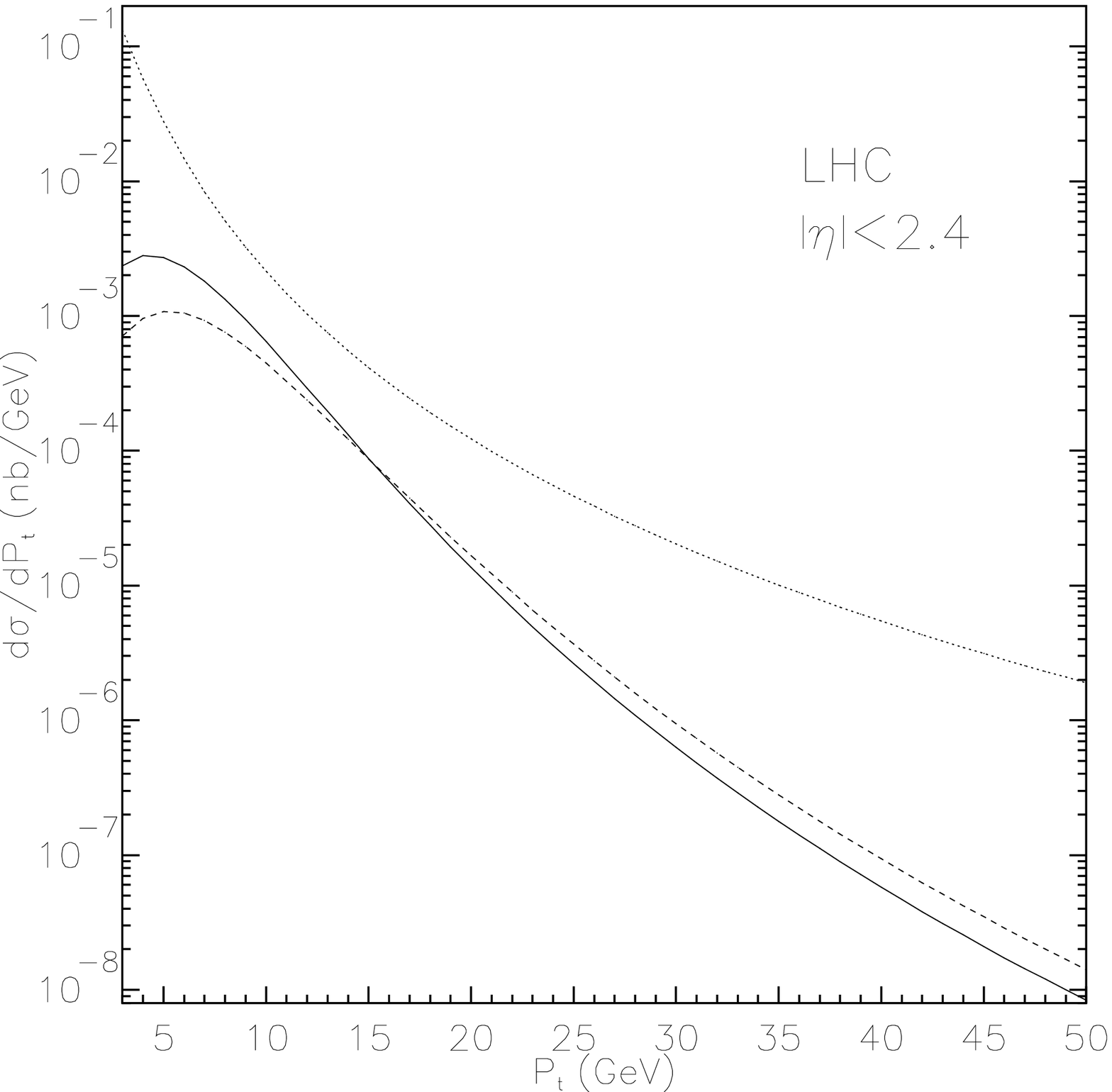,width=8cm}\\
\end{tabular}
\caption{The pair production of $\Upsilon$ (solid line) and $\eta_b$ (dashed line)
in the CSM at the hadron
colliders. The dotted line corresponds to the pair production of $\Upsilon$ that
come from the gluon fragmentation
process in the COM.}
\label{fig:Upsilon-Etab}
\end{figure}

The $p_t$ distributions of $\Upsilon$ and $\eta_b$ production are
shown in Fig.~\ref{fig:Upsilon-Etab}. The $p_t$ distribution is enhanced by
an order or more in magnitude at the LHC than that at the Tevatron. But unlike the case of $J/\psi$
pair production, the pair production of $\Upsilon$ in the COM dominate over
that in the CSM in the whole $p_t$ region, and even is two order or more in magnitude larger
than that in the CSM at large $p_t$.

\begin{figure}
\begin{tabular}{cc}
\epsfig{file=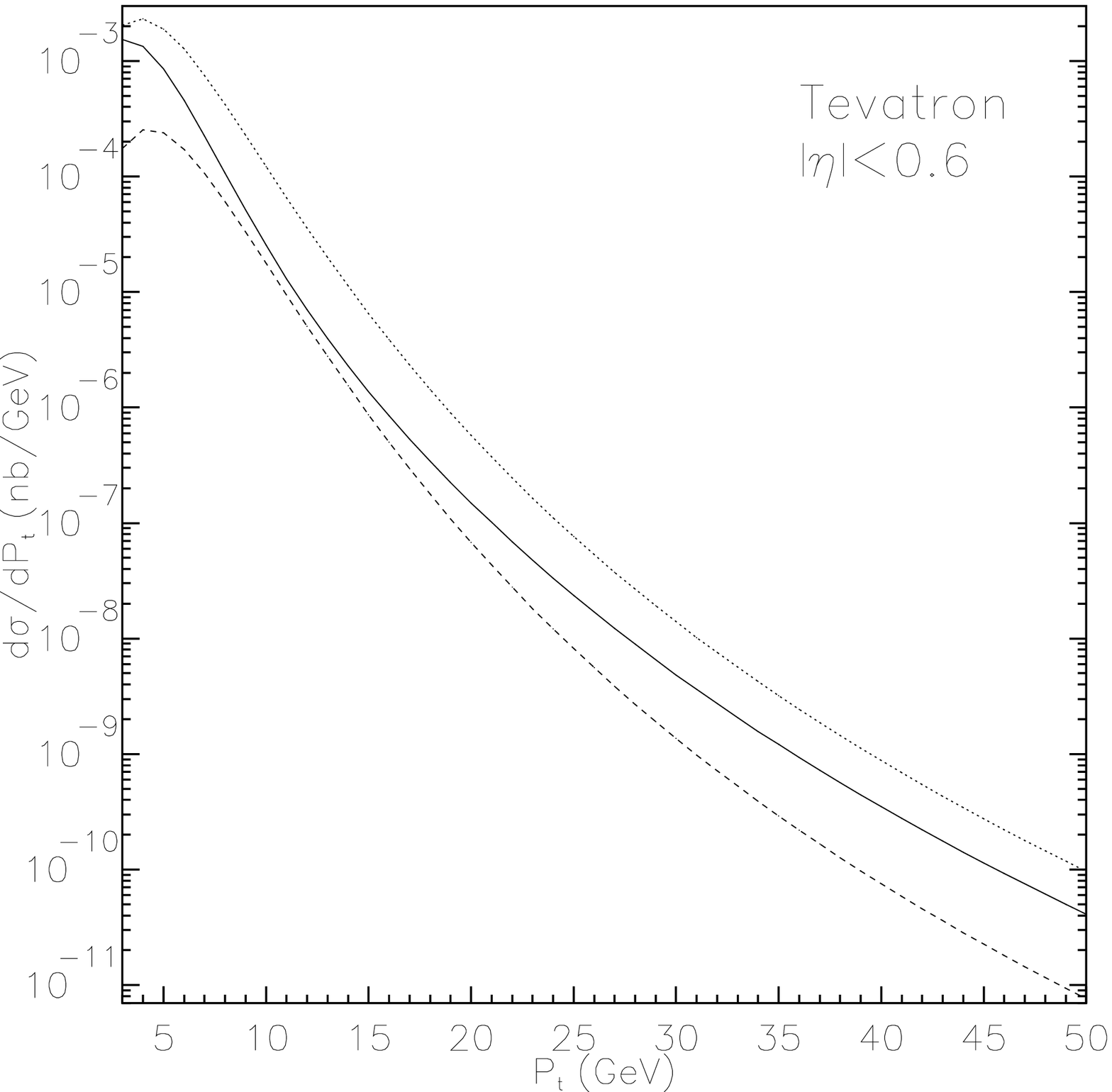,width=8cm}&
\epsfig{file=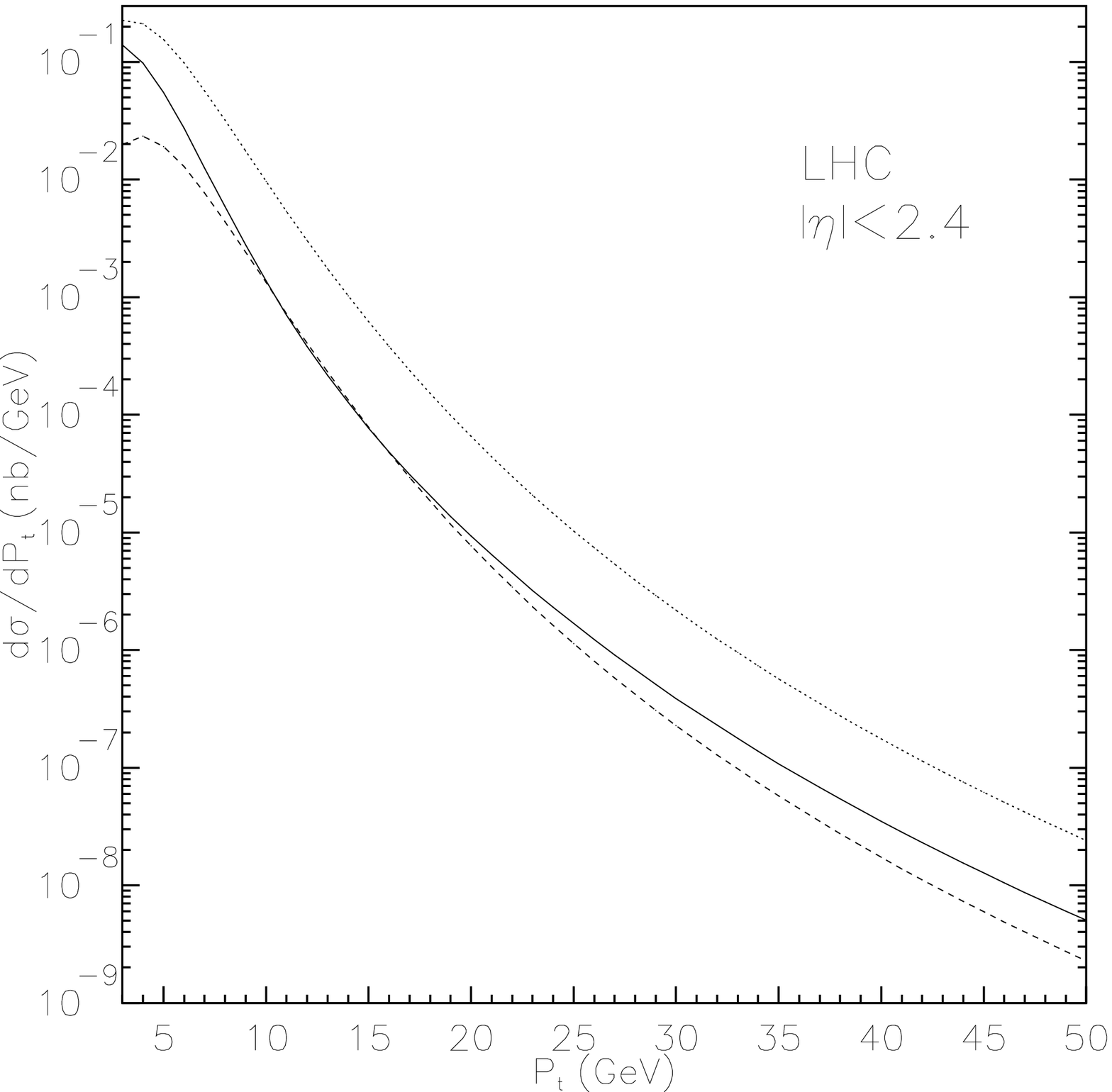,width=8cm}\\
\end{tabular}
\caption{The production of $B_c \bar{B_c}$ (solid line),~$B_c \bar{B_c^*}$ (dashed line) and
$B_c^* \bar{B_c^*}$ (dotted line) in the CSM at the Tevatron and LHC.}
\label{fig:Double-Bc}
\end{figure}

Because the $B_c$ and $B_c^*$ are consist of quarks with different
flavors, there is no contribution from gluon fragmentation processes
in the COM. So in Fig.~\ref{fig:Double-Bc}, we only give the $p_t$
distribution of the pair production for $B_c\bar{B_c}$,
$B_c\bar{B_c^*}$ and $B_c^*\bar{B_c^*}$ in the CSM. The pair
production of $B_c^*\bar{B_c^*}$ is dominant in the whole $p_t$
region. There are little difference between the production of the
three final states with double-heavy flavor mesons in the CSM.

From the Table I and the figures, it can be seen
that all of the cross sections are enhanced when the center-of-mass
energy is increased. This is because that with the fixed
$p_t$, the larger the $\sqrt{s}$ is, the
smaller the momentum fraction $x$ of the parton is. In small $x$ region,
the parton distribution function of gluon increases rapidly.

\section{Summary}

In this paper, we have investigated the leading order pair
production of S-wave heavy quarkonium at hadron colliders in the
color-singlet mechanism (CSM) and estimated the contributions from
the gluon fragmentation process in the color-octet mechanism (COM)
for comparison. With the matrix elements extracted previously in
leading order calculations, the numerical results show that the
production rates are quite large for the pair production processes
at the LHC. The $p_t$ distribution of double $J/\psi$ production in
the CSM is dominant over that in the COM when $p_t$ is smaller than
about 8 GeV. For the production of double $\Upsilon$, the
contribution of the COM is always larger than that in the CSM. There
are large differences in the theoretical predictions between the CSM
and COM for the $p_t$ distributions in the large $p_t$ region, and
this is useful in clarifying the effects of COM on the quarkonium
production. Furthermore, since to produce a pair of quarkoniums with
different $C$-parity is forbidden in the CSM at the leading-order,
the observation of these processes could be a positive support for
the COM.  We also investigate the pair productions of S-wave $B_c$
and $B_c^*$ mesons, and the measurement of these processes may be
useful to test the CSM and extract the LDMEs for the $B_c$ and
$B_c^*$ mesons.

After our work was completed \cite{Li:2008phd}, a paper appeared
\cite{Qiao:2009kg}, in which Qiao, Sun, and Sun calculated the
double $J/\psi$ production at the LHC. They focused on the
polarizations of the double $J/\psi$. We focused on the cross
sections of double heavy quarkonia $J/\psi$, $\eta_c$, $\Upsilon$,
$\eta_b$, as well as the double heavy flavored $B_c$, and $B_c^*$
mesons. Both the two papers discuss the test of the COM. Our color
octet and color singlet double $J/\psi$ cross sections are
consistent with their result \cite{Li:2008phd}.

\begin{acknowledgments}
We thank Prof. Jian-Xiong Wang and Dr. Ce Meng for helpful
discussions. This work was supported by the National Natural Science
Foundation of China (No 10675003, No 10721063, No 10805002).
\end{acknowledgments}

\section*{Appendix}

In this appendix we give the parton level cross section for the pair
production of $\eta_c(\eta_b)$ and $J/\psi(\Upsilon)$ respectively.
Here m is the mass of the corresponding meson and the s, t, u are
the Mandelstam variables defined as
\begin{eqnarray}
\label{Mand4}
s &=& (k_1+k_2)^2, \nonumber \\
t &=& (k_1-P)^2, \nonumber \\
u &=& (k_2-P)^2,
\end{eqnarray}
where $k_1$, $k_2$ and $P$ are the momenta of the initial gluons and
one of the final quarkonium states.

1.~$g+g \to \eta_c+\eta_c$
\begin{eqnarray}
&&\hspace{-0.8cm}\frac{d\hat{\sigma}}{dt}=\frac{{\alpha_s}^4 \pi
|R(0)|^4}{162 m^2 s^8 (m^2-t)^4 t^2
   (-2 m^2+s+t)^2 (-m^2+s+t)^4} (331776 t^2
m^{28}-239616 t^2 (7 s\nonumber \\ &&\hspace{0.4cm} +18 t)
m^{26}+256 t (27 s^3+14257 t s^2+81792 t^2 s +101412
   t^3) m^{24}-64 t (543 s^4  +72538 t
s^3\nonumber \\ &&\hspace{0.4cm}+687780 t^2 s^2+1876968 t^3 s
+1498176 t^4) m^{22} +4 (9 s^6  +18144 t
   s^5+18144 t   s^5
   \nonumber \\
&&\hspace{0.4cm}+1018456 t^2 s^4+13409824 t^3 s^3 +60363088 t^4 s^2
+104825088 t^5 s +60673536 t^6) m^{20}\nonumber \\ &&\hspace{0.4cm}
   -4 (45 s^7+20364 t
   s^6 +693676 t^2 s^5+10769920 t^3 s^4 +70075680 t^4 s^3+199182848 t^5 s^2\nonumber \\ &&\hspace{0.4cm}
    +248330880 t^6 s+111310848 t^7) m^{18} +(369
   s^8+53568 t s^7+1532448 t^2 s^6 +24574336 t^3 s^5\nonumber \\ &&\hspace{0.4cm}+210621248 t^4\hspace{-0.04cm} s^4
   +870605312 t^5\hspace{-0.04cm} s^3+1761630976 t^6\hspace{-0.04cm} s^2 +1686196224 t^7
   \hspace{-0.04cm}s  +610384896 t^8) m^{16}\nonumber \\ &&\hspace{0.4cm} -4 (99 s^9
   +5496 t s^8+172056 t^2 s^7 +2626424 t^3 s^6+27129120 t^4 s^5
    +152150816 t^5
   s^4 \nonumber \\ &&\hspace{0.4cm} +446139360 t^6 s^3 +687430400 t^7 s^2+527406336 t^8 s+158754816 t^9) m^{14}
   +2 (117 s^{10}\nonumber \\ &&\hspace{0.4cm}
   +3420 t\hspace{-0.04cm}  s^9+131584 t^2\hspace{-0.04cm}
   s^8+1843176 t^3\hspace{-0.04cm}  s^7+20064012 t^4\hspace{-0.04cm}  s^6+143086496 t^5\hspace{-0.04cm}  s^5 +571404544 t^6\hspace{-0.04cm}  s^4\nonumber \\ &&\hspace{0.4cm}
   +1267269888 t^7 s^3+1553349056 t^8 s^2+984517632 t^9
   s +251817984 t^{10}) m^{12}-4 (18 s^{11}\nonumber \\ &&\hspace{0.4cm}+588 t s^{10}+21495 t^2 s^9
   +304486 t^3 s^8 +2927138 t^4 s^7+23491008 t^5
   s^6+120913176 t^6 s^5\nonumber \\ &&\hspace{0.4cm}+363365760 t^7 s^4 +636265120 t^8 s^3
   +639890816 t^9 s^2+342648576 t^{10} s+75727872 t^{11})
   m^{10}\nonumber \\ &&\hspace{0.4cm}+(9 s^{12}+810 t s^{11}+20676 t^2 s^{10}+360708 t^3 s^9+3111538 t^4 s^8
   +23098608 t^5 s^7\nonumber \\ &&\hspace{0.4cm}+138861808 t^6\hspace{-0.04cm}
   s^6+537887488 t^7\hspace{-0.04cm} s^5+1266810688 t^8\hspace{-0.04cm} s^4+1806816000 t^9\hspace{-0.04cm} s^3+1525860352 t^{10}\hspace{-0.04cm} s^2\nonumber \\ &&\hspace{0.4cm}
   +702443520 t^{11} s+135945216 t^{12})
   m^8-4 t (45 s^{12}+903 t s^{11}+18636 t^2 s^{10}+181253 t^3 s^9 \nonumber \\ &&\hspace{0.4cm}
   +1173996 t^4 s^8+7001216 t^5 s^7+32407688 t^6 s^6+98050752
   t^7 s^5+186725360 t^8 s^4 \nonumber \\ &&\hspace{0.4cm}+222162048 t^9\hspace{-0.04cm} s^3
   +160423104 t^{10}\hspace{-0.04cm} s^2
   +64384128 t^{11}\hspace{-0.04cm} s+11031552 t^{12}) m^6+2 t (9
   s^{13}+396 t \hspace{-0.04cm}  s^{12}\nonumber \\ &&\hspace{0.4cm}
   +4962 t^2 s^{11}+58165 t^3 s^{10}+384536 t^4 s^9
   +2056988 t^5 s^8+10083672 t^6 s^7+37041746 t^7 s^6\nonumber \\ &&\hspace{0.4cm}+90302176
   t^8 s^5+142532816 t^9 s^4+144160704 t^{10} s^3+90317984 t^{11} s^2 +31965696 t^{12} s
   \nonumber \\ &&\hspace{0.4cm}
   +4893696 t^{13}) m^4-4 t^2 (45
   s^{13}+312 t s^{12}+2799 t^2 s^{11}+20392 t^3 s^{10}+103737 t^4 s^9 \nonumber \\ &&\hspace{0.4cm}
   +480982 t^5 s^8+1974841 t^6 s^7+5914228 t^7 s^6 +11914204
   t^8 s^5+15898752 t^9 s^4 \nonumber \\ &&\hspace{0.4cm}+13880896 t^{10} s^3+7636608 t^{11} s^2
   +2406528 t^{12} s+331776 t^{13}) m^2+t^2 (s+t)^2 (18
   s^{12}\nonumber \\ &&\hspace{0.4cm}
   +90 t s^{11}+447 t^2 s^{10}+3162 t^3 s^9+14485 t^4 s^8
   +57520 t^5 s^7+241296 t^6 s^6+755200 t^7 s^5\nonumber \\ &&\hspace{0.4cm}+1481344 t^8
   s^4+1774080 t^9 s^3+1267200 t^{10} s^2+497664 t^{11} s +82944 t^{12}))
\end{eqnarray}

2.~$g+g \to J/\psi+J/\psi$
\begin{eqnarray}
\frac{d\hat{\sigma}}{dt}&=&\frac{16 {\alpha_s}^4 \pi |R(0)|^4}{81 m^2 s^8
(m^2-t)^4
   (-m^2+s+t)^4} (7776 m^{24}-432 (73
s+216 t) m^{22}\nonumber \\ &&+6 (9085 s^2+60336 t s+85536 t^2)
m^{20}-16
   (3629 s^3+37686 t s^2+117855 t^2 s\nonumber \\ &&+106920 t^3) m^{18}
   +4 (11927 s^4+151588 t s^3+745674 t^2 s^2+1470960 t^3
   s\nonumber \\ &&+962280 t^4) m^{16}-4 (7761 s^5+109608 t s^4
   +699467 t^2 s^3+2173908 t^3 s^2\nonumber \\ &&+3055320 t^4 s+1539648 t^5)
   m^{14}+2 (6952 s^6+117893 t s^5+897043 t^2 s^4\nonumber \\ &&
   +3741980 t^3 s^3+8278410 t^4 s^2+8872416 t^5 s+3592512 t^6) m^{12}-2
   (1899 s^7\nonumber \\ &&+43398 t s^6+405618 t^2 s^5+2113568 t^3 s^4
   +6394090 t^4 s^3+10762584 t^5 s^2\nonumber \\ &&+9189936 t^6 s+3079296 t^7)
   m^{10}+(587 s^8+19710 t s^7+244772 t^2 s^6\nonumber \\ &&+1603468 t^3 s^5
   +6229962 t^4 s^4+14478304 t^5 s^3+19359816 t^6 s^2+13582080 t^7
   s\nonumber \\ &&+3849120 t^8) m^8-2 (20 s^9+1185 t s^8+22153 t^2 s^7
   +193780 t^3 s^6+965358 t^4 s^5\nonumber \\ &&+2928368 t^5 s^4+5431786 t^6
   s^3+5949528 t^7 s^2+3508920 t^8 s+855360 t^9) m^6\nonumber \\ &&+(s^{10}
   +76 t s^9+3756 t^2 s^8+52062 t^3 s^7+353472 t^4
   s^6+1398834 t^5 s^5\nonumber \\ &&+3421754 t^6 s^4+5210968 t^7 s^3
   +4784622 t^8 s^2+2414880 t^9 s+513216 t^{10}) m^4\nonumber \\ &&-4 t^2 (s+t)^2
   (9 s^7+649 t s^6+6460 t^2 s^5+29630 t^3 s^4+74435 t^4 s^3+105156 t^5 s^2
   \nonumber \\ &&+77868 t^6 s+23328 t^7) m^2+2 t^4 (s+t)^4
   (349 s^4+2304 t s^3+6192 t^2 s^2\nonumber \\ &&+7776 t^3 s+3888 t^4))
\end{eqnarray}

\newpage

\end{document}